\RequirePackage{silence}
\documentclass[article,10pt]{elsarticle}
\usepackage{lineno,hyperref}
\modulolinenumbers[5]
\journal{Astroparticle Physics Journal}
\usepackage{graphicx}
\usepackage{amsmath}

\usepackage{amsfonts}
\usepackage{float}
\usepackage{bm}
\usepackage[perpage,symbol*]{footmisc}
\usepackage{ae,aecompl}
\usepackage{array}
\usepackage{soul}
\usepackage{mathtools}
\usepackage{multirow}
\usepackage{xcolor}
\usepackage{caption,subcaption}
\usepackage[margin=0.7in]{geometry}
\newcommand{\appropto}{\mathrel{\vcenter{
		\offinterlineskip\halign{\hfil$##$\cr 
			\propto\cr\noalign{\kern2pt}\sim\cr\noalign{\kern-2pt}}}}}

\begin{document}

\begin{frontmatter}

\title{Identification of Gamma Ray Pulsar Candidates in the \emph{Fermi}-LAT 4FGL-DR4 Unassociated Sources Using Supervised Machine Learning}

\author[mymainaddress1,mysecondaryaddress]{A. Pathania\corref{mycorrespondingauthor1}}
\cortext[mycorrespondingauthor1]{Corresponding author}
\ead{apathania@barc.gov.in}
\author[mymainaddress1,mysecondaryaddress]{K. K. Singh\corref{mycorrespondingauthor2}}
\cortext[mycorrespondingauthor2]{Corresponding author}
\ead{kksastro@barc.gov.in}
\author[mymainaddress2]{S. K. Singh}
\author[mymainaddress1,mysecondaryaddress]{A. Tolamatti}
\author[mymainaddress3]{B. B. Singh}
\author[mymainaddress1,mysecondaryaddress]{K. K. Yadav}
\address[mymainaddress1]{Astrophysical Sciences Division, Bhabha Atomic Research Centre, Mumbai - 400 085, India}
\address[mysecondaryaddress]{Homi Bhabha National Institute, Anushaktinagar, Mumbai - 400 094, India}
\address[mymainaddress2]{Division of Remote Handling \& Robotics, Bhabha Atomic Research Centre, Mumbai - 400 085, India}
\address[mymainaddress3]{Department of High Energy Physics, Tata Institute of Fundamental Research, Mumbai - 400005, India}
\begin{abstract}

	The Large Area Telescope (LAT) on board the \emph{Fermi} Gamma-ray Space Telescope has been continuously providing good quality 
	survey data of the entire sky in the high energy range from 30 MeV to 500 GeV and above since August 2008. A succession of gamma-ray 
	source catalogs is published after a comprehensive analysis of the \emph{Fermi}--LAT data. The most recent release of data in the 
	fourth \emph{Fermi}--LAT catalog of gamma-ray sources (4FGL-DR4), based on the first 14 years of observations in the energy band 
	50 MeV-1 TeV, contains 7195 sources. A large fraction ($\sim$ 33\%) of this population has no known counterparts in the lower wave 
	bands. Such high energy gamma-ray sources are referred to as unassociated or unidentified. An appropriate classification of these 
	objects into known type of gamma-ray sources such as the active galactic nuclei or pulsars is essential for population studies and 
	pointed multi-wavelength observations to probe the radiative processes. In this work, we perform a detailed classification of the 
        unassociated sources reported in the 4FGL-DR4 catalog using two supervised machine learning techniques-Random Forest and 
	Extreme Gradient Boosting. We mainly focus on the identification of new gamma-ray pulsar candidates by making use of different 
	observational features derived from the long-term observations with the \emph{Fermi}--LAT and reported in the incremental 4FGL-DR4 catalog. 
	We also explore the effects of data balancing approach on the classification of the \emph{Fermi}--LAT unassociated sources. 
\end{abstract}
\end{frontmatter}

\section{Introduction}
\label{Introduction}
The launch of \emph{Fermi} Gamma-ray Space Telescope in August 2008 has revolutionised the field of high energy astrophysics in a very short 
time span of less than two decades \citep{Atwood2009,Ajello2021}. Over the time, the number of high energy gamma-ray sources and their types 
has increased many fold and several successive catalogs have been published based on the \emph{Fermi}-data \citep{Abdo2009,Abdo2010,Nolan2012,Acero2015,
Abdollahi2020,Ballet2020,Abdollahi2022}. All the \emph{Fermi} gamma-ray source catalogs (FGL), published so far and named 0 to 4FGL in succession, are 
dominated by only two class of sources called Active Galactic Nuclei and Pulsars. The number of unassociated sources (without possible counterparts 
in other wave bands) has also successively increased. The most recent data release of the \emph{Fermi}-Collaboration (4FGL-DR4), using 14 years of the 
survey data of entire sky during the period August 4, 2008 to August 2, 2022 with Large Area Telescope (LAT) in the energy range 50 MeV-1 TeV, reports 
7195 sources \citep{Ballet2023}. Of the 7195 sources in the 4FGL-DR4 catalog, 2428 are found to be unassociated i.e. their nature remains unknown 
in any other energy or frequency regime. The most dominant population in the successive \emph{Fermi}-LAT catalogs is represented by the extragalactic 
sources mainly active galactic nuclei or blazars in particular. The second dominant class of sources in the \emph{Fermi}-LAT catalogs is shared by 
the pulsars mostly populated in the Milky Way Galaxy. 
\par
\begin{figure*}	
		\begin{center}
		\includegraphics[width=0.60\columnwidth,angle=0]{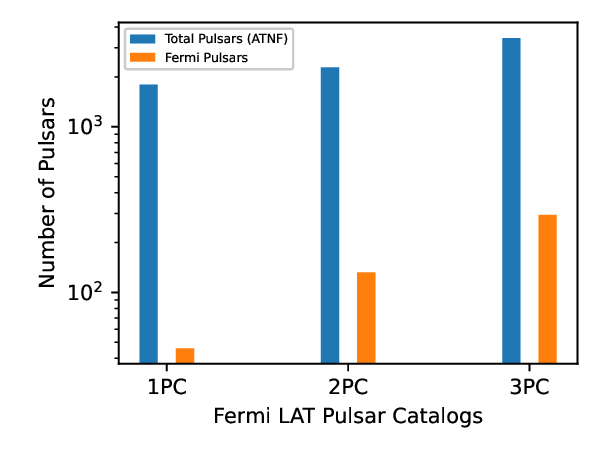}
		\caption{Evolution of the population of pulsars in the contemporaneous \emph{Fermi}-LAT and ATNF catalogs.}
		\label{atnf-lat}
		\end{center}	
\end{figure*}

Pulsars are rapidly rotating strongly magnetized neutron stars (surface magnetic field in the range 10$^8$ to 10$^{14}$ G), which emit pulsed non-thermal 
radiation across the whole electromagnetic spectrum \citep{Gold1968}. The pulse periods are measured over a wide range from milliseconds to a few 
seconds \citep{Reddy2022,Pathania2023}. According to their pulse or rotation period, pulsars are classified as millisecond pulsars and 
normal/canonical pulsars \citep{Bhattacharyya2022}. Millisecond pulsars having rotaion periods less than 30 ms, also known as recycled pulsars, are 
believed to be old sources with weaker magnetic field. Normal or non-recycled pulsars, exhibiting the strongest magnetic fields, have pulse 
periods $\ge$ 30 ms and are relatively young. The pulsars are mainly detected in the radio observations by searching for the radio 
scintillations \citep{Bruzewski2021} and the first pulsar was discovered in 1967 during a radio astronomy project \citep{Hewish1968}. Since 
this discovery, more than 3500 pulsars have been detected so far mostly in radio observations as per the catalog regularly published by the 
Australia Telescope National Facility (ATNF)\footnote{https://www.atnf.csiro.au/research/pulsar/psrcat/}. The most recent third pulsar catalog (3PC), 
based on 12 years of high energy gamma-ray observations with the \emph{Fermi}-LAT, reports about 340 gamma-ray pulsars and 
candidates \citep{Smith2023}. This represents a very small fraction ($<$ 10\%) of the known pulsar population. A comparison of the contemporary population 
of pulsars detected by radio observations and reported in the ATNF catalog with the gamma-ray pulsars reported in the successive \emph{Fermi}-LAT pulsar 
catalogs is shown in Figure \ref{atnf-lat}. Only a few pulsars are observed at optical wavelengths as their magnetospheres are weak optical emitters and 
a few dozens are observed in the X-ray band \citep{Manchester2005}. The population of known gamma-ray pulsars is almost equally 
dominated by the millisecond pulsars and young pulsars \citep{Caraveo2014,Frail2018,Smith2023}. This marks the most significant contribution of the 
\emph{Fermi}-LAT in the field of pulsar astronomy as only five gamma-ray pulsars were identified (by the EGRET \citep{Hartman1999}) prior to its launch in 2008. 
The light curves of gamma-ray pulsars are steady and their energy spectra, as measured by the \emph{Fermi}-LAT, exhibit 
curvature. It requires addition of an exponential cutoff at few GeV energies \citep{Ballet2023,Pathania2023}. Above this energy, the spectrum decreases 
very rapidly, making it difficult to detect the pulsars at GeV-TeV energies with the ground-based gamma-ray telescopes. Consequently, only a few 
pulsars ($\sim$ 5) have been detected in the very high energy regime \citep{Terzi2021,Pathania2023}. Highly regular pulsation or light curve of pulsars 
allows stacking of signal from different epochs and increases detection probability in both radio and gamma-ray wavebands \citep{Smith2023}. 
\par
Being located mostly in the Galactic plane, pulsars are wonderful objects and powerful tools to probe different aspects of physics and astrophysics 
including radiative processes under strong electromagnetic field \citep{Romani2014}, stellar evolution \citep{Lorimer2008}, dark matter \citep{Desai2016} 
and fundamental physics \citep{Desai2024}. Pulsar observations are also used as potential tools in different astrophysical experiments such as pulsar timing 
array for low frequency gravitational wave detection \citep{Kramer2013}, equation of state of neutron stars in nuclear astrophysics \citep{Lattimer2004}, tests 
of the general theory of relativity \citep{Berti2015} etc. Therefore, search for pulsars is important for statistical study of pulsar population, their physical 
properties, classification and probing the fundamental physics. However, timing observations are computationally expensive for pulsar search as narrow pulses need 
to be resolved with very high time resolution. Machine learning and artificial intelligence based algorithms are extensively used to identify the \emph{Fermi-LAT} 
unassociated sources according to their observational features in gamma-rays or other wave-bands \citep{Ackermann2012,Doert2014,Salvetti2017,Yi2017,Kang2019,Kova2020,
Kaur2023,Tolamatti2023,Parkinson2016}. The \emph{Fermi}-LAT detected sources are characterized in terms of the spectral characteristics and positional accuracy. 
Mirabal et al. (2016) took the advantage of the spectral measurements, reported in the third \emph{Fermi}-LAT catalog (3FGL) \citep{Acero2015}, to identify the pulsar 
candidates within the 3FGL unassociated sources by using supervised machine learning classifiers and found that 34 additional candidates can be accommodated by existing 
pulsar population synthesis models \citep{Mirabal2016}. Finke et al. (2021) employed a fully connected deep neural network and a recurrent neural network to 
classify the unassociated sources reported in the 4FGL-DR2 catalog \citep{Ballet2020} into  Active Galactic Nuclei and Pulsars classes using photon energy 
spectrum and time series measurements from the \emph{Fermi}-LAT as inputs \citep{Finke2021}. Germani et al. (2021) used an ensemble of artificial neural networks 
to characterise the 4FGL-DR2 unassociated sources based on the likelihood of being a pulsar or subclasses of blazars \citep{Germani2021}. 
Balakrishnan et al. (2021) proposed a semi-supervised generative adversarial network with better classification performance for identification of pulsar 
candidates among the 4FGL-DR2 unassociated sources \citep{Balakrishnan2021}. In the present work, we use supervised machine learning (Random Forest and 
Extreme Gradient Boosting) to classify the 4FGL-DR4 unassociated sources into pulsars and blazars with a focus on the identification of gamma-ray pulsar 
candidates. The paper is organized as following. 
In Section \ref{SML}, we briefly describe the supervised machine learning technique. The data-set used in this work is described in 
Section \ref{data-set}. We discuss the results in detail in Section \ref{results}. Finally, we conclude this study in Section \ref{conclusions}.

\section{Supervised Machine Learning}\label{SML}
Artificial Intelligence based methods including machine learning are emerging as common place for value added higher level data products 
for scientific research in different domains of detection, classification and regression \citep{Fluke2020}. It deals with solving a given problem 
by training or learning through the available information. Machine Learning refers to an approach which learns from the input data using computer 
algorithms and models to make predictions. In the supervised learning \citep{Baron2019}, a predictive model is built through the mapping between 
an input space and a known output or the ground truth. It describes a very complex non-linear relationship between input and output variables and 
therefore completely differs from the model fitting wherein the model is predefined. The relationship or model derived from the supervised learning 
is said to be approximately correct if the predicted output lies within an acceptable margin of error from the ground truth. Prediction of correct 
result over a large sample has a high probability for a correct model. Models in the supervised machine learning generally perform classification 
and regression. This study deals with the classification and aims to classify a large population of latest unidentified gamma-ray sources into 
Active Galactic Nuclei and Pulsars classes. For a binary classification problem, the prediction outcomes are divided into two classes: Positive and Negative. 
In the present work, Pulsars belong to the positive class whereas Active Galactic Nuclei constitute the negative class. Therefore, four types of outcomes 
can occur during the classification of unassociated sources: True Positive (TP), True Negative (TN), False Positive (FP), and False Negative (FN). 
TP means a source is predicted as pulsar and it belongs to the Pulsar class. TN implies that a source is classified as active galactic nuclei and it belongs to the 
Active Galactic Nuclei class. FP indicates that an active galactic nuclei is predicted to be a pulsar. FN occurs when a pulsar is classified as an active galactic 
by the machine learning model. 
\par
The supervised learning algorithms are generally applied in two stages: training and testing. The entire data set is randomly divided into training and 
testing samples in a typical fraction of 75\% and 25\% respectively \citep{Choi2020}. In the first stage, the training data is used to learn input to 
output mapping, i.e. obtaining an appropriate model for solving the problem at hand. In the next stage, the testing data set is used to assess the 
model output against the ground truth, i.e. determining the validity of the derived model. At this stage, the performance and quality of a machine 
learning model or trained model is evaluated in terms of various parameters like \emph{accuracy, precision, recall, F1 score} and \emph{receiver 
operating characteristic (ROC) curve}, based on the different combinations of TP, TN, FP and FN. For a balanced train data set, the accuracy of a 
binary classifier is defined as
\begin{equation}\label{eqn-accuracy}
	\mathrm{Accuracy~=~\frac{Correct~Predictions}{Total~Predictions}~=~\frac{TP + TN}{TP + TN + FP + FN}}
\end{equation} 
It is a measure of how accurately a supervised machine learning algorithm is able to classify the sources in their respective classes. The above definition 
can give misleading results if the train data set is unbalanced. In case of unbalanced train data set, a new metric called \emph{balanced accuracy} is 
estimated as \citep{Urbanowicz2015}
\begin{equation}\label{eqn-accuracy-unbal}
    \mathrm{Balanced~ Accuracy = 0.5 \times \left[\frac{TP}{TP +FN}+\frac{TN}{TN + FP}\right]}
\end{equation}
for a binary classifier. The first and second terms in the right hand side of Equation \ref{eqn-accuracy-unbal} are referred to as sensitivity and 
specificity for each class respectively. Balanced Accuracy measures the accuracy of a model by giving equal weightage to both majority and 
minority classes. Precision is defined as
\begin{equation}\label{eqn-precision}
	\mathrm{Precision = \frac{TP}{TP + FP}}
\end{equation}
It is an important metric to minimize the misclassification of negative classs sources as positive class by a trained model. 
Recall is defined as 
\begin{equation}\label{eqn-recall}
	\mathrm{Recall = \frac{TP}{TP + FN}}
\end{equation}
It is a measure of the ability of a binary classifier how correctly it classifies a positive class source as positive class. Recall is also known as 
Sensitivity or True Positive Rate (ratio of number of true positive predictions to the total number of positive events in the input data) of a model. A high recall rate 
underlines the fact that the model is able to successfully predict majority of positive class outputs. On the other hand, a high specificity or True Negative 
Rate (ratio of number of true negative predictions to the total number of negative events in the input data) indicates that the model 
is successfully extracting most of the negative class sources from the data. F1 score, the harmonic mean of precision and recall,  is estimated as 
\begin{equation}\label{eqn-F1score}
    \mathrm{F1~Score = \frac{2\times Precision \times Recall}{Precision + Recall}}
\end{equation}
It combines the precision and recall of a model into a single metric and is primarily evaluated for the performance evaluation of two classifiers. 
The ROC curve is a graphical tool to evaluate the performance of a classifier. It is produced by plotting the sensitivity or true positive rate 
against (1-specificity) or false positive rate. In the best-case scenario, the true positive rate and the false positive rate are 1 and 0 respectively. 
In the worst-case scenario, the true positive rate is linearly proportional to the false positive rate. The area under ROC curve is used to define 
an optimal model or classifier. A larger value of area under ROC curve signifies higher accuracy of the classifier.\\
\par
In this work, we explore the potentials of two well known supervised learning classifiers Random Forest \citep{Breiman2001} and Extreme Gradient 
Boosting \citep{Friedman2001} for classification of a sample of \emph{Fermi}-LAT unassociated sources into the  Active Galactic Nuclei and Pulsars classes. 
Random Forest (RF) is based on the ensemble learning methods. It consists of a number of de-correlated or independent decision trees with random 
inputs (through bootstrapping) from the training data set and each tree casts a vote on the ouput. Votes from all independent trees are counted 
in the classification. The majority vote is considered as the final result leading to the building of random forests. The class of the object is 
decided by the collective decision of the entire forest. However, in case of Extreme Gradient Boosting (XGB), which also contains a group of 
decision trees, all decision trees are not independent. It works by employing the decision trees in a serial manner with each tree learning from 
the mistake of previous tree by using a gradient boosting parameter called learning rate. Higher the learning rate, stronger is the corrections 
implemented in the model building. The typical value of learning rate is $\sim 0.1$. 
\begin{figure*}	
	        \begin{center}
		\includegraphics[width=0.60\columnwidth,angle=0]{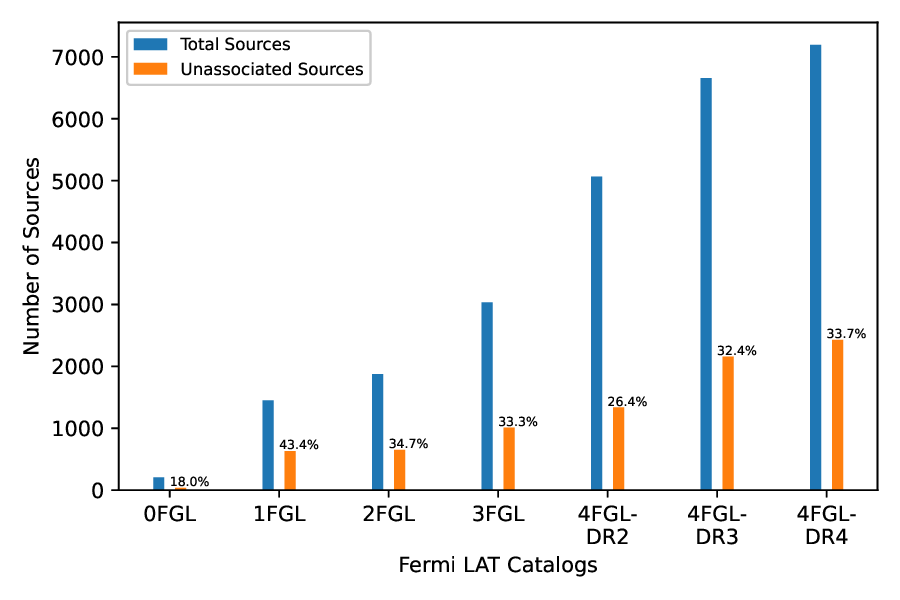}
		\caption{Comparison of the population of unassociated sources with that of the total number of gamma-ray sources reported in the successive 
				\emph{Fermi}-LAT catalogs.}
		\label{unass-puls}		
		\end{center}			
\end{figure*}
\begin{table}
\caption{Summary of the 4FGL-DR4 Source Classes. A detailed description of these classes can be found in the 4FGL-DR3/DR4 catalogs \citep{Abdollahi2022,Ballet2023}.}
\begin{tabular}{|p{2.5cm} p{3.5cm} p{6.5cm} p{3cm}|}
\hline
Class & Identified/Associated & Included Sub-class (as per `CLASS1' tag) & Number of Sources\\
\hline
Active Galactic Nuclei 	&Yes & bcu, BCU, bll, BLL, fsrq, FSRQ, rdg, RDG, nlsy1, NLSY1, agn, AGN, ssrq, sey \& css &4016 \\
Pulsar 	&Yes & PSR, psr, MSP \& msp & 320\\
Other 	&Yes & SNR, PWN, GC, Binary systems, normal galaxies \& other than above listed subclasses & 431\\
Unassociated & No & - & 2428\\
 \hline
\end{tabular}
\label{tab:datatype}
\end{table}

\section{Data Set and Processing}\label{data-set}
The number of LAT sources increases steadily as the all-sky survey continues, with the unassociated sources remaining roughly a third 
of the total (Figure \ref{unass-puls}). In the latest 4FGL-DR4 catalog 
(version $\textit{gll\_psc\_v34.fit}$), 7195 gamma-ray sources are reported \citep{Ballet2023}. All these sources have been further divided 
into 24 subclasses and are designated as identified or associated on the basis of correlated timing signature at different wavelenghts or positional 
coincidence \citep{Parkinson2016}. In the present work, we have grouped these sources into 4 classes viz, Active Galactic Nuclei, Pulsar, Other, and 
Unassociated as described in Table \ref{tab:datatype}. It can be clearly seen that majority of the sources belong to the Active Galactic Nuclei 
class with a share of $\sim$ 55.8\%, and the second dominant population of 
known sources belongs to the Pulsar class with a fraction of $\sim$ 4.4\%. However, a significant fraction $\sim$ 33.7\% remains unassociated. 
Constraints related to the incremental identification of newly added sources in the 4FGL-DR4 catalog with respect to the 3FGL catalog 
indicate that $\sim$ 5\% of the unassociated sources are likely to be pulsars. The 4FGL-DR4 catalog provides informations related to the 
position (source coordinates), gamma-ray spectrum (parameters corresponding to different spectral shapes and their uncertainties), flux (differential flux 
and energy flux in 8 energy bands starting from 50 MeV to 1 TeV and their test statistics (TS) values), timing (yearly flux history, variability index, 
fractional variability), significances and associations (identification/association tags, TeV-catalog flag, association with recent \emph{Fermi}-LAT catalogs) 
under more than 150 columns (including sub-columns) for all the sources. 
\subsection{Data Selection}\label{data-selection}
Since the pulsar population is mostly concentrated in the low galactic latitude region \cite{Pathania2023}, we have not considered 
the positional information as an input feature for the classification of a source as this is not a physical parameter to decide the nature of source and 
may lead to a bias in the predictions. We also exclude the best fit spectral informations of sources from the input data set. Rather, we use all the 
spectral parameters and their uncertainties corresponding to different spectral forms like Power law (PL), Log-Parabola (LP) and Power-Law with 
super/sub-Exponential Cutoff (PLEC). We use the derived quantities like Fractional Variability, Variability index etc. instead of the flux history information 
for temporal properties of the sources. Observational constraints require that the pulsar candidates should have PLEC spectrum, a spectral 
cutoff energy and low Variability index ($<$ 24.5) as pulsars are considered to be steady sources. The spectral cutoff energy for PLEC spectrum 
is defined as \citep{Abdollahi2022,Pathania2023} 
\begin{equation}\label{eqn-cutoff-ene}
\mathrm{ E_c^{PLEC} = E_0 \times \left(\frac{b^2}{d}\right)^{1/b}}
\end{equation}
where $\rm E_0$ is the pivot energy (given in the catalog), $\rm b$ and $\rm d$ are exponential index and spectral curvature respectively. 
Another important feature is the Hardness Ratio ($\rm {HR_{ij}}$), which can be estimated as \citep{Ackermann2012} 
\begin{equation}\label{eqn-HR}
\mathrm{	{ HR_{ij} = \frac{F_j - F_i}{F_j + F_i}}}
\end{equation}
where $\rm F_i$ and $\rm F_j$ are energy fluxes in $\rm i^{th}$ and $\rm j^{th}$ energy bands. $\mathrm {HR_{ij}}$ values for five energy bands 100 MeV - 300 MeV, 
300 MeV - 1 GeV, 1 GeV - 3 GeV, 3 GeV - 10 GeV, and 10 GeV - 30 GeV, are used as input features. We have deliberately ignored the extreme energy 
ends 30 MeV - 100 MeV and 30 GeV - 1 TeV having wide point spread function and low photons statistics respectively, leading to relatively poor 
sensitivity of the \emph{Fermi}-LAT in these bands. The value of $\mathrm{HR_{ij}}$ lies between -1 and +1. We have also ignored few 
informations like `Unc\_PLEC\_Exp\_Index', `PLEC\_Epeak', `Unc\_PLEC\_Epeak' etc for which more than 30\% entries are empty. Due to the above filtrations, 
the features available under more than 150 columns in the 4FGL-DR4 catalog reduce to 85. We perform the Kolmogorov-Smirnov (K-S) test on active galactic nuclei and pulsar 
populations to determine the most prominent features (which differ significantly) and use the Kendall-$\tau$ correlation analysis to eliminate the highly 
correlated features. The KS-test statistical value of greater than 0.35 and $|\tau|> 0.75$ together significantly reduce the total number of features 
from 85 to 18. Among these 18 features, we perform $\rm log$ transformations of a few which are either highly skewed or have extremely low values 
like $\rm Signif\_Avg$, $\rm Variability$\_$Index$, and $\rm Flux_{1000}$ etc.

\begin{figure*}	
	        \begin{center}
		\includegraphics[width=0.60\columnwidth,angle=0]{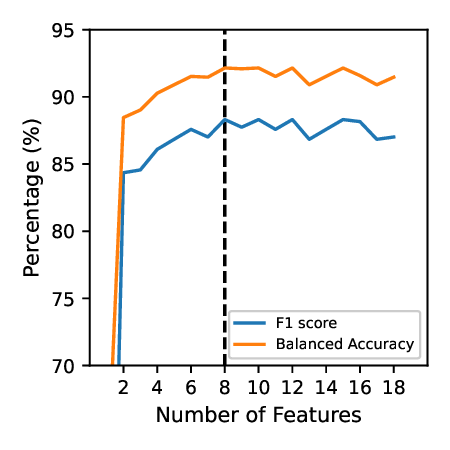}
		\caption{Balanced Accuracy and F1-score as a function number of features.}
		\label{fig:selected-features}		
		\end{center}			
\end{figure*}
\subsection{Feature Selection}\label{feature-selection}
\begin{figure*}	
	        \begin{center}
		\includegraphics[width=0.80\columnwidth,angle=0]{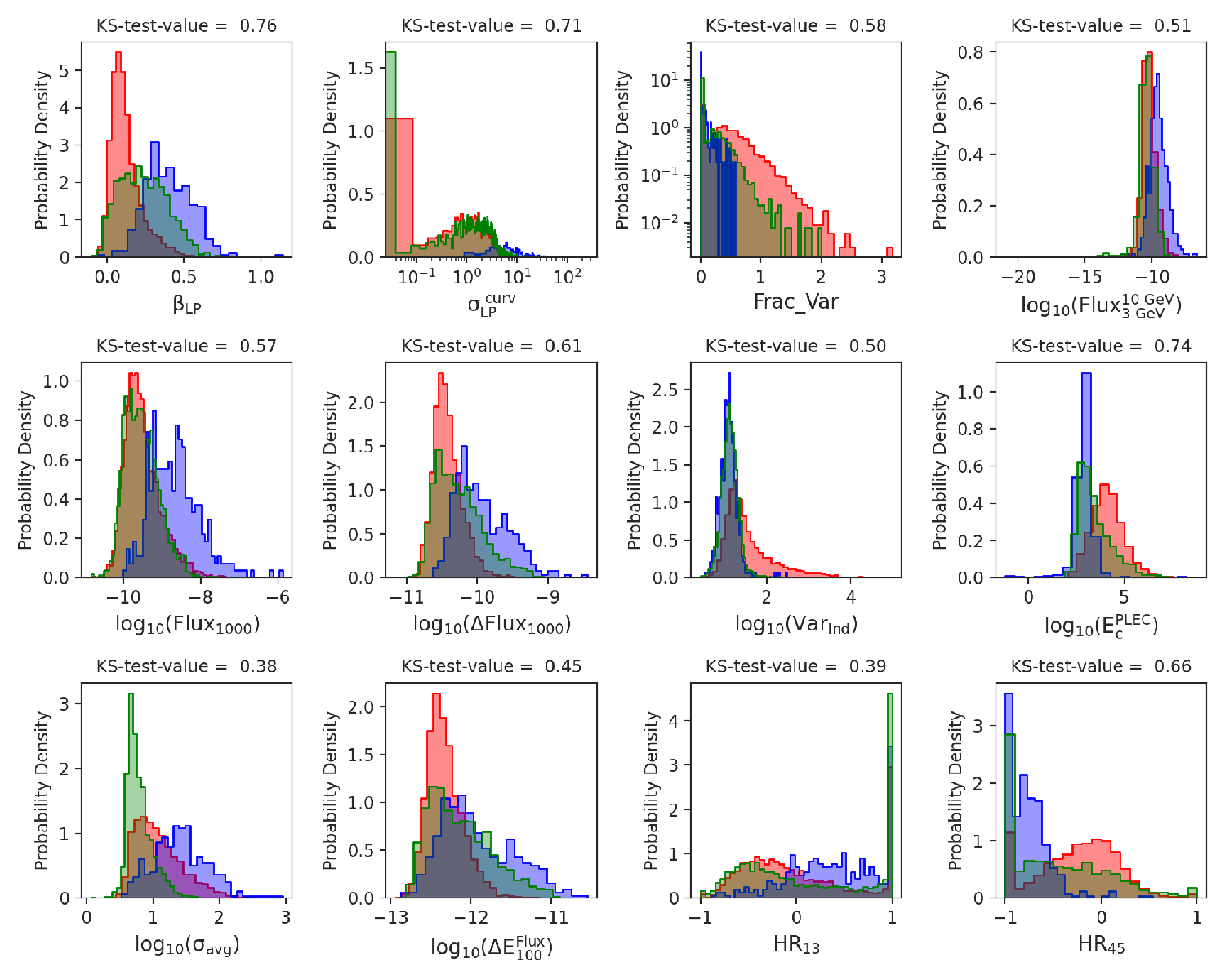}
		\caption{Distributions of 12 features selected for Pulsar (Blue), Active Galactic Nuclei (Red) and Unassociated source (Green) classes  
			with corresponding K-S test values (for Pulsar and Active Galactic Nuclei) on the top of each plot.}
		\label{fig:feature}		
		\end{center}			
\end{figure*}

The process by which an effective set of features can be created from the pool of total available features for better performance and 
reduced computational complexity is referred to as the Feature Selection. The data selection procedure, explained in Section \ref{data-selection}, 
helps to significantly narrow down the required information useful for the Active Galactic Nuclei and Pulsar binary classification. Recent works 
on the problem of classification of sources report that better results can be obtained using supervised machine learning algorithms with 
10 or less input features \citep{Luo2020,Zhu2024}. Luo et al. (2020) have shown that an accuracy of better than 95\% can be achieved using only 
5 input features under the framework of Recursive Feature Elimination (RFE) method \citep{Luo2020}. In this work, we also employ 
RFE framework \citep{Richert2013} for completing the task of feature selection among the total 18 available feautres on the basis of balanced 
accuracy. In the RFE framework, a backward selection method, an effective set of features is selected by eliminating the least important 
feature iteratively. This is done through a classification method which generates importance score for each feature. Intitally, entire pool 
of 18 features is used to evaluate the performance of RF classifier along with the RFE framework and in each iteration the least important 
feature is eliminated successively down to 1 feature. Balanced Accuracy and F1-score, evaluated in each iteration, 
are used to find the effective set of features. In the present work, we have used the enitre dataset consisting of 320 pulsars and 3768 active galactic 
nuclei for the purpose of feature selection. This dataset is further randomly divided into Train and Test datasets in the ratio of 75\% and 25\% 
respectively. The variation of estimated Balanced Accuracy and F1-score as a function of the number of features is shown in Figure \ref{fig:selected-features}. 
It is clearly observed that both Balanced Accuracy and F1-score have maximum values for 8 input features. However, same performance can be acheived for 
12 features also. These 12 selected features are listed in Table \ref{tab:ip-features}. Only first 8 features (Table \ref{tab:ip-features}, Sr. No. 1 - 8) 
are used as final set of prominent features for classification of unassociated sources using RF and XGB models. The distributions of selected input 
features are presented in Figure \ref{fig:feature} for known pulsars, active galactic nuclei, and unassociated sources. It is evident that the 
distributions of all the parameters for the two classes: Pulsar (blue) and Active Galactic Nuclei (red) are very well separated from each other. 
This is also supported by the corresponding K-S test statistics values given on the top of each distribution. However, the distributions of unassociated 
source class (green) overlap with those of  Pulsar and Active Galactic Nuclei. This indicates that majority of unassociated sources should have association 
with either of the class. We designate Pulsar and Active Galactic Nuclei classes with 1 and 0 respectively for broad classification of unassociated sources 
as potential pulsar or active galactic nuclei candidates in the present work. 

\begin{table}
	\caption{Summary of 12 input features selected for classification. Only first 8 features consitute the prominent set of input 
	features for binary classification in the present work. A detailed description of these parameters can be found in \citep{Abdollahi2022}.}
\begin{tabular}{|p{.5cm} p{13cm} p{3cm}|}
\hline
Sr. No. &Input Feature &Symbol\\
\hline
1  &Curvature Parameter (when fitting with LP)  	&$\mathrm{\beta_{LP}}$ \\
2  &Significance of fit improvement between PL and LP   &$\mathrm{\sigma_{LP}^{curv}}$\\
3  &Fractional variability computed from the fluxes each year &Frac\_Var  \\
4  &Logarithm of photon flux from 3 GeV to 10 GeV &$\mathrm{log_{10} ( Flux_{3GeV}^{10GeV})}$ \\
5  &Logarithm of integral photon flux in the energy range 1 GeV to 100 GeV &$\mathrm{log_{10} (Flux_{1000})}$  \\
6  &Logarithm of $1\sigma$ error in the integral photon flux in the energy range 1 GeV to 100 GeV & $\mathrm{log_{10} (\Delta Flux_{1000})}$  \\
7  &Logarithm of sum of 2$\times$ log(Likelihood) difference between the flux fitted in each time interval and the average flux over the full 
	catalog interval &$\mathrm{log_{10} (Var_{Ind})}$  \\
8  &Logarithm of Cut-off energy as defined in Equation \ref{eqn-cutoff-ene} & $\mathrm{log_{10} (E_c^{PLEC})}$ \\
9  &Logarithm of source significance in units of $\sigma$ in the energy range 100 MeV to 1 TeV &$\mathrm{log_{10} (\sigma_{avg})}$ \\
10 &Logarithm of $1\sigma$ error on energy flux from 0.1 GeV to 100 GeV & $\mathrm{log_{10} (\Delta E_{100}^{Flux})}$  \\
11 &Hardness ratio for the energy bands (0.1 - 0.3 GeV) and (1 - 3 GeV) &$\mathrm{HR_{13}}$ \\
12 &Hardness ratio for the energy bands (3 - 10 GeV) and (10 - 30 GeV) & $\mathrm{HR_{45}}$ \\

\hline
\end{tabular}
\label{tab:ip-features}
\end{table}
\subsection{Training and Testing}\label{data-process}
\begin{figure*}	
	        \begin{center}
		\includegraphics[width=0.80\columnwidth,angle=0]{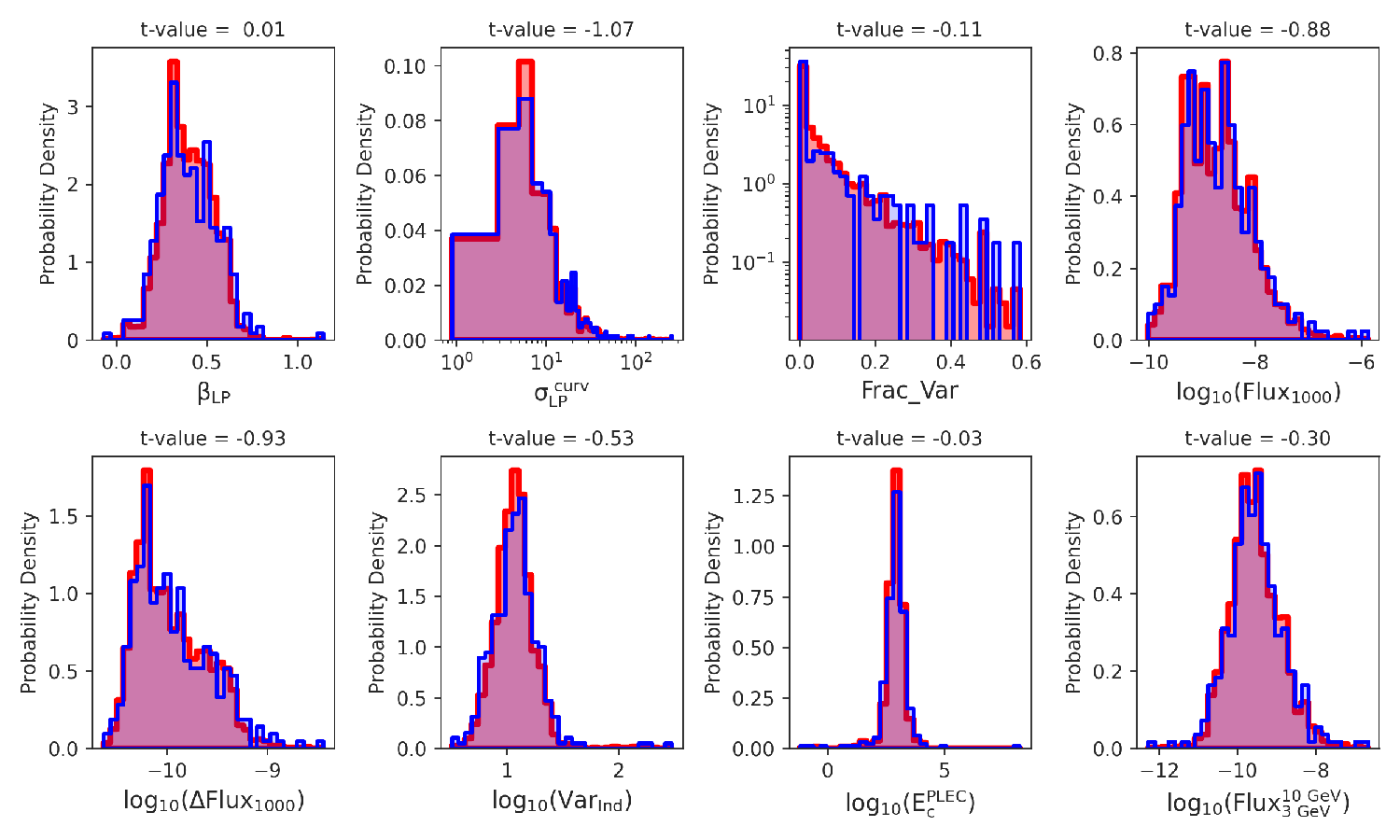}
		\caption{Distributions of various input features of the pulsars in the Training dataset A (Blue) and Training dataset B (Red) along with the 
			corresponding values of the t-statistics on the top.}
		\label{fig:smote-data}		
		\end{center}			
\end{figure*}

The final dataset, consisting of 8 known features for 320 pulsars and 3768 active galactic nuclei, is generated to classify 2257 unassociated sources. 
The data is randomly divided into training and testing datasets in the ratio of 75\% and 25\% respectively. Therefore, the training data 
(defined as Training dataset A) comprises observed features of 241 pulsars and 2825 active galactic nuclei, whereas 79  pulsars and 943 
active galactic nuclei constitute the testing data. It is important here to mention that the two classes, Pulsar and Active Galactic Nuclei, are not 
equally represented in the Training dataset A since the number of pulsars is less than that of the active galactic nuclei. Such imbalance in the 
training data may lead to a bias in favour of the dominant class or against the under-represented class for the  ensemble-based classifiers \citep{Last2017}. 
Therefore, we employ Synthetic Minority Over-sampling Technique (SMOTE) to balance the representation of Pulsar class in the training data. 
SMOTE follows k-nearest neighbor approach to generate the synthetic samples by randomly selecting the data point from the minority class and 
interpolating between the existing data points \citep{Chawla2002,Chawla2011}. We refer this augmented data as the Training dataset B wherein 
the imbalance between Pulsar and Active Galactic Nuclei classes has been addressed through SMOTE. The distributions of input selection features 
for pulsars in the Training dataset A and B are presented in Figure \ref{fig:smote-data}. It is observed that the distributions of all the input 
features, generated using SMOTE (Training dataset B), match very well with those of the actual data (Training dataset A). In order to further quantify 
the level of similarity between the two distributions, we have performed the statistical Student's t-test. The t-values estimated for distributions 
of each of the input features in the Training datasets A and B (given on the top of corresponding plots) are close to zero. This suggests that the 
distributions of all the parameters in both the datasets A and B resemble each other very well. 
\par
We first optimize the hyper-parameters using \textit{GridSearchCV}\footnote{https://scikit-learn.org/stable/modules/generated/
sklearn.model$\_$selection.GridSearchCV.html} for RF and XGB classifiers based on the Training datasets A and B. The datasets are divided into 
5 cross-validation sets for both the classifiers. In case of Training dataset A, the best hyper-parameters for RF are found 
to be $\rm n\_ estimator = 80$ and $\rm max\_ depth = 7$, whereas for XGB, the best hyper-parameters are  $\rm n\_ estimator = 130$, 
$\rm max\_ depth = 4$ with $\rm learning\_ rate = 0.08$. However, the best hyper-parameters are obtained as $\rm n\_ estimator = 600$, 
$\rm max\_ depth = 7$ and $\rm n\_ estimator = 500$, $\rm max\_ depth = 4$ for RF and XGB respectively  with $\rm learning\_ rate = 0.1$ in 
case of Training dataset B. We train the model for  the Training dataset A and B using the corresponding hyper-parameters derived above. 
The average accuracy of both the classifiers is evaluated using 
\textit{stratified-shuffle-cross-validation} \footnote{https://scikit-learn.org/stable/modules/generated/sklearn.model$\_$selection.StratifiedShuffleSplit.html} 
by dividing the Training datasets A and B into 5 equal size subsets. It is found to be better than 97\% for both the training datasets.

\section {Results and Discussion}\label{results}
We use the above RF and XGB-based binary classification models to classify more than 2200 unassociated sources in the latest 4FGL-DR4 catalog 
into Pulsar and Active Galactic Nuclei classes. The performance of the classifiers, importance of different input features in the classification, and 
results from the classification of unassociated sources are described below.
\subsection{Performance of Classifiers}\label{performance}
We estimate various assessment metrics as defined in Section \ref{SML} to evaluate the classifier performance using the test data. 
The values of these metrics for RF and XGB classifiers trained with datasets A and B are listed in Table \ref{tab:metrics}. 
The model accuracy of RF and XGB are found to be better than 97.5\% for both the datasets. However, the effect of data balancing 
using SMOTE (dataset B) can be clearly seen in the balanced accuracy of the classifiers with a difference of $\sim$ 8\% 
with respect to the original data (dataset A). For XGB classifier, the balanced accuracy is close to 99\% in case of dataset B. 
A significant improvement in the Recall-value is also noticed for both the classifiers with dataset B. A Recall-value greater 
than 98\% implies a better classification of true pulsar sources as a potential pulsar candidate by both RF and XGB classifiers. 
A slight decrease in the value of specificity is found for RF classifier trained with the dataset B. Similarly, the precision of 
RF classifier also decreases by $\sim$ 11\% when balanced data (dataset B) is used for training. The F1 score shows a 
significant improvement of $\sim$ 8\% for XGB with training using balanced data (dataset B). Therefore, the performance 
of XGB classifier with balanced data training is found to be the best for classification of the unassociated sources in 
the 4FGL-DR4 catalog.
\begin{table}
\caption{Assessment metrics for performance evaluation of RF and XGB models trained with original (dataset A) and balanced data (dataset B).}
\begin{center}
\begin{tabular}{c c c c c c c c c}
\hline
Classifier &Training Dataset &Accuracy &Balanced Accuracy &Specificity  &Precision  &Recall  &F1 score\\
	   &                 &(\%)	&(\%)		  &(\%)		&(\%)	    &(\%)     &(\%)\\ 
\hline
RF  &A &97.74 &89.50 &99.25 &90.00 &79.74 &84.56   \\
RF  &B &97.84 &98.25 &97.77 &78.78 &98.73 &87.64   \\
XGB &A &98.14 &92.61 &99.15 &89.47 &86.07 &87.74  \\
XGB &B &99.21 &99.57 &99.15 &90.80 &100.0 &95.18  \\
\hline
\end{tabular}
\label{tab:metrics}
\end{center}
\end{table}
\subsection{Feature Importance}\label{importance}
\begin{figure*}	
	        \begin{center}
		\includegraphics[width=0.7\columnwidth,angle=0]{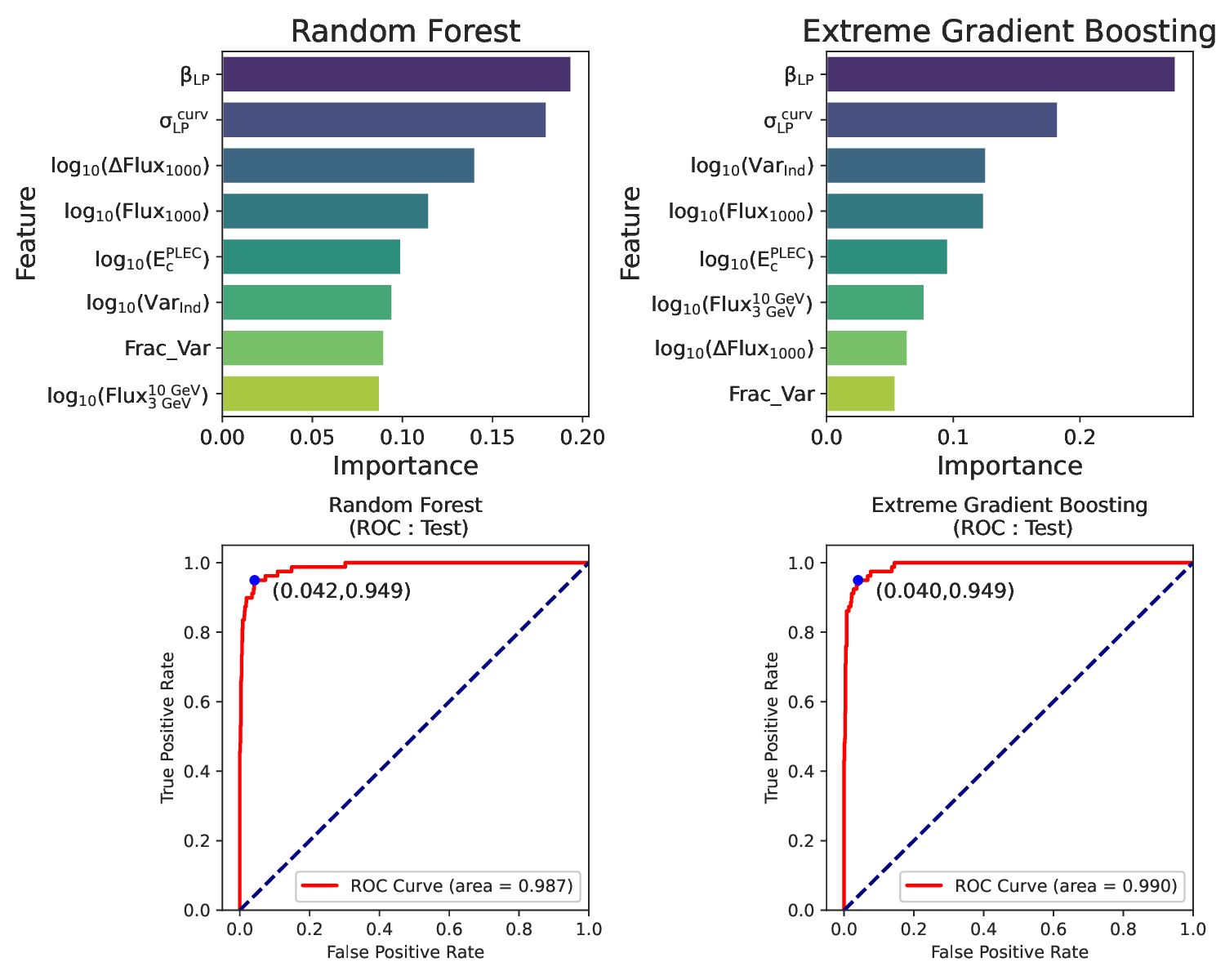}
		\caption{Importance of various input features and ROC curve for RF and XGB classifiers trained with original dataset A.}
		\label{fig:feature-A}		
		\end{center}			
\end{figure*}

\begin{figure*}	
	        \begin{center}
		\includegraphics[width=0.7\columnwidth,angle=0]{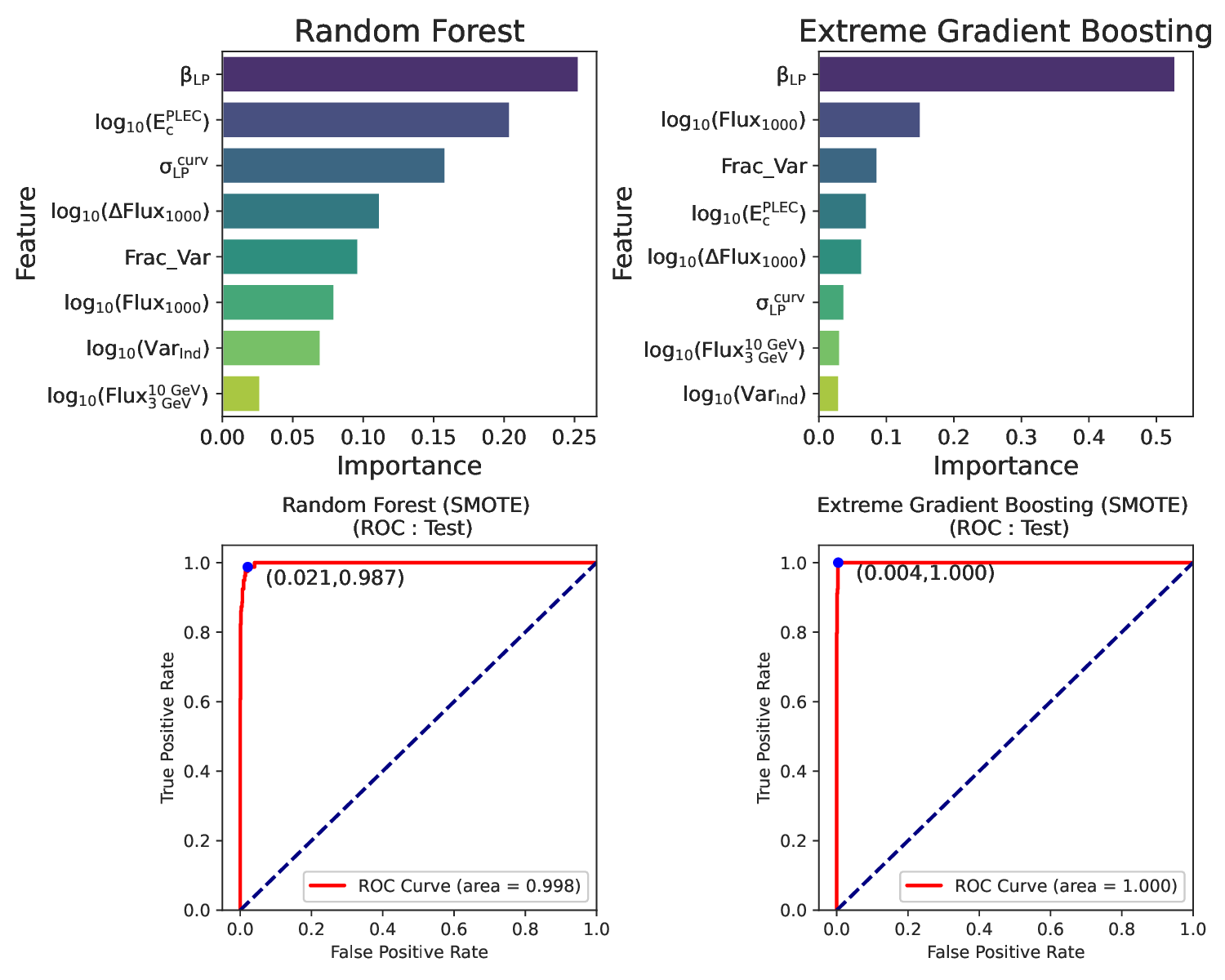}
			\caption{Same as Figure \ref{fig:feature-A} for balanced dataset B (SMOTE).}
		\label{fig:feature-B}		
		\end{center}			
\end{figure*}

The feature importance is estimated to quantify the strength of a particular input feature in deciding the class by a given classifier based on the 
machine learning algorithms. It also helps in associating the corresponding feature as a signature of the whole class. The histograms of the importance 
of 8 input features along with the ROC curves and value of area under the curve for RF and XGB classifiers corresponding to the training 
datasets A and B are shown in Figure \ref{fig:feature-A} and \ref{fig:feature-B} respectively. It is observed that use of  balanced data (dataset B) 
during the training process helps in increasing the value of area under ROC curve for both RF and XGB classifiers. This implies a significant improvement 
in the performance of two binary classifiers. The values of area under ROC curve as unity and F1 score as $\sim$ 95\% suggest that the XGB model trained 
with dataset B is expected to perform better than RF. The most important input feature for the binary classification is the curvature parameter 
($\beta_{LP}$) in case of RF and XGB models (Figure \ref{fig:feature-B}). $\mathrm log_{10} (Flux_{1000})$ and $\mathrm log_{10} (Var\_Ind)$ turn out 
to be the second and third most important features for deciding the class of unassociated sources using XGB classifier. As pulsars are observed to be 
relatively stable source with respect to the active galactic nuclei, importance of Var\_Ind is justified while considering the long-term temporal 
behavior of the sources. Thus spectral parameters ($\mathrm \beta_{LP}$ and $\mathrm E_c^{PLEC}$) and temporal behavior play crucial role in the 
classification of unassociated \emph{Fermi}-LAT sources. 
\subsection{Classification of Unassociated Sources}\label{classification}
\begin{figure*}	
	        \begin{center}
		\includegraphics[width=0.80\columnwidth,angle=0]{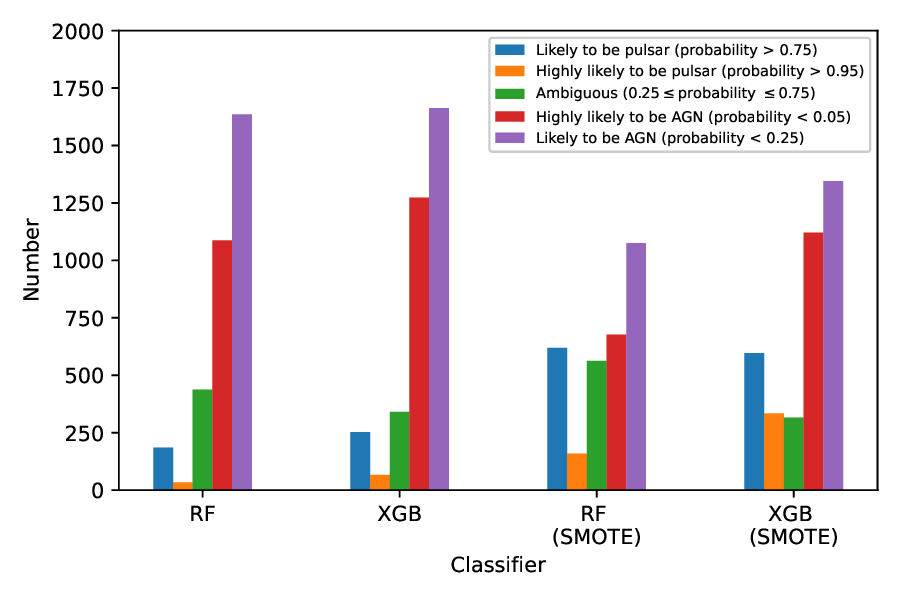}
		\caption{Distribution of number of probable Pulsar and Active Galactic Nuclei candidates from the classification of 2257 unassociated 
			sources by RF and XGB classifiers using the training datasets A (original) and B (SMOTE).}
		\label{fig:classification}		
		\end{center}			
\end{figure*}

The RF and XGB models, trained using the training datasets A and B, are employed to classify 2257 unassociated sources of the 4FGL-DR4 catalog into 
Pulsar and Active Galactic Nuclei classes. Sources classified as class 1 with an association probability $> 75\%$ and $> 95\%$ are referred to as 
likely Pulsar and highly likely  Pulsar candidates respectively. An association probability $< 25\%$ and $< 5\%$ characterizes the sources as likely 
and highly likely Active Galactic Nuclei (class 0) candidates respectively. However, sources with association probability $\geq 25\%$ but $\leq 75\%$ 
are termed as ambiguous. Based on these criteria, the final classification results from this work are presented in Figure \ref{fig:classification}. 
The RF classifier with training dataset A is able to identify 185 Pulsar and 1635 Active Galactic Nuclei candidates, whereas a significant 
population of 437 unassociated sources remains ambiguous. Among the probable 185 Pulsar candidates, 34 are highly likely to be potential Pulsar candidates, 
and 1087 out of 1635 Active Galactic Nuclei candidates are highly likely to be active galactic nuclei. Similarly, the XGB classifier, based on training 
dataset A classifies 253 Pulsar, 1663 Active Galactic Nuclei and 341 ambiguous candidates. Among them, 67 and 1273 are highly likely Pulsar and 
Active Galactic Nuclei candidates respectively. 
The number of predicted Pulsar candidates significantly increases when both the classifiers, trained on the balanced data (dataset B), 
are used for classification of 2257 unassociated sources. In this, RF (SMOTE) and XGB (SMOTE) successfully identify 619 and 596 Pulsar candidates respectively. 
Accordingly, the numbers of highly likely to be Pulsar candidates increase to 159 and 334 for RF and XGB classifiers respectively. 
It is interesting to note that number of ambiguous sources slighlty decreases from 341 to 316 for XGB while using the balanced data for training. 
On the contrary, the population of ambiguous sources increases from 437 to 563 in case of RF with balanced data training. The number of highly likely to be 
Active Galactic Nuclei candidates does not change significantly for XGB but a drastic decrease from 1075 to 677 is noted in case of RF trained with 
balanced data. Thus, XGB classifier with original (imbalanced) and balanced data training predicts more highly likely Pulsar 
candidates among 2257 unassociated sources of the 4FGL-DR4 catalog than RF classifier. 
Results obtained in this work represent an incremental classification of latest population of the \emph{Fermi}-LAT unassociated sources with respect 
to the recent classification by Zhu et al. 2024 \citep{Zhu2024} wherein multiple machine learning methods excluding XGB have been used to 
classify 2157 unassociated sources in the 4FGL-DR3 catalog. 
A complete list of unassociated sources, classified in the present work, is provided as the supplementary material.   
  
\subsection{Covariate Shift Analysis}\label{covariate}
The basic assumption of a supervised machine learning algorithm is that training (known sources) and testing (unknown sources) datasets should 
follow same distribution. If the distributions of training and testing datasets are not identical, a machine learning algorithm will suffer from sample 
selection bias. This is also known as covariate shift or the Malmquist bias. Presence of covariate shift can affect the performance of a machine learning 
algorithm, when employed on an unknown dataset. The issue of covariate shift is very common and natural in astronomy as bright sources are identified 
easily \& dominate the training dataset with respect to the faint sources which are not identified easily and dominate the testing or unknown 
dataset \citep{Richards2011,Richards2012}. In case of high energy gamma-ray observations, distribution of unassociated sources may not resemble the 
distributions of detected active galactic nuclei and pulsars since unassociated sources exhibit relatively low statistical signficance (faint objects) 
with respect to the already identified/associated sources. We have performed a covariate shift analysis to ensure the performance and robustness of  
RF and XGB classifiers in the present study.
\par
The 4FGL-DR4 catalog also provides informations which can be used to verify the class of a gamma-ray source in previous 3FGL (`ASSOC\_FGL' column) 
or 4FGL-DR3 (`ASSOC\_4FGL' column) catalogs. In 3FGL catalog, total numbers of pulsars and active galactic nuclei were 245 and 1938 
respectively. However, the numbers of pulsars and active galactic nuclei increased to 316 and 3602 respectively in 4FGL-DR3 catalog. 
This indicates that $\sim$ 5\% of the newly identified sources from 3FGL to 4FGL-DR3 catalog are pulsars. These newly identified sources 
(71 pulsars and 1664 active galactic nuclei) in 4FGL-DR3 catalog as compared to 3FGL catalog can be used to check the robustness of the 
classifiers and effect of including weak features (Table \ref{tab:ip-features}, Sr. No. 9 - 12) on binary classification. For this purpose, 
we create two dataset named C and E using 3FGL and 4FGL-DR3 catalogs respectively. The dataset C consists of identified active galactic nuclei 
and pulsars tagged with 3FGL whereas dataset E consists of newly identified active galactic nuclei and pulsars in 4FGL-DR3. As dataset C is an 
imbalanced dataset, data balancing is performed using SMOTE as described earlier to generate a balanced training dataset D. 
The efficiency of both RF and XGB classifiers is evaluated using training datasets C and D followed by test dataset E under the covariate shift.

\subsubsection{Effect of Weak Features}\label{covariate-weak}
We use the training dataset C and test dataset E to check the effect of weak features (Sr. No. 9 - 12 in Table \ref{tab:ip-features}) 
for RF classifier and the results are summarized in Table \ref{tab:weak-feature}. The balanced accuracy for dataset E is obtained to be 65.67\% 
and 64.41\% corresponding to set of 8 and 12 input features. However, the numbers of true pulsar classification and negative active galactic nuclei 
are higher and lower respectively for 8 features with respect to 12 features. This implies that inclusion of weak features (in case of 12) do not improve the classifier 
performace. Therefore, a model with less number of features (8) is an optimal choice for the binary classification beacuse of its relatively 
simple structure and less computational cost.
\begin{table}
	\caption{Results from the assessment of effect of weak features on the binary classification of sources in dataset E (71 pulsars and 1664 
	active galactic nuclei) using RF classifier. N1 and N2 respectively represent the number of sources identified as highly likely  
	(probability $>$ 95\%) 	and likely (probability $>$ 75\%) of the given class.}
\begin{center}
\begin{tabular}{c c c c c}
\hline
Features Included &True Pulsar &True Active Galactic Nuclei &False Pulsar &False Active Galactic Nuclei\\
		  &(N1,N2)     &(N1,N2)                     &(N1,N2)      &(N1,N2)\\ 
\hline
8  	&6, 24  &1484, 1623 &0, 5 &7, 22   \\
12 	&6, 22  &1505, 1628 &0, 5 &11, 26 \\
\hline
\end{tabular}
\label{tab:weak-feature}
\end{center}
\end{table}
\subsubsection{Actual Performance of Classifiers}\label{actual-performance}
We use 8 prominent input features to examine the actual performance of RF and XGB classifiers under covariate shift for the test 
dataset E (71 pulsars and 1664 active galactic nuclei). The binary classification results are summarized in Table \ref{tab:covariate-E}. 
As the dataeset E consists of newly identified sources (pulsars and active galactic nuclei), it can be considered as the 
more realistic test data for performance evaluation of the classifiers with respect to the test datasets derived from A and B, which are 
dominated by the relatively bright sources. The performance evaluation using the dataset E suggests that the balanced accuracy of both the 
classifiers reduces from $\sim$ 90\% (for the test datasets derived from A and B) to $\sim$ 65\% under the covariate shift. This reduction 
in the  balanced accuracy is due to the selection bias effects as the testing dataset E consists of relatively faint sources. 
Also, training with the balanced dataset D increases the true and false positive rates with later being the dominant for both the classifiers 
under the covariate shift. This artificially boosts the number of probable pulsar candidates. However training with the imbalanced dataset C 
gives a realistic estimate of the number of probable pulsar candidates.
A comparison of RF and XGB models suggest that later outperforms the former in identifying the highly likely class of Pulsars for imbalanced dataset C. 
Recall of XGB is better than RF while predicting Pulsar class sources, but precision of RF is better than XGB. In case of balanced dataset D, RF turns 
out to be an effective classifier for the likely to be Pulsar class with similar recall and better precision than XGB.
\par
Different combinations of individually trained RF and XGB classifiers can also be used to identify pulsar candidates in the test dataset E to 
examine their collective robustness. Results from 8 such combinations of RF and XGB having better true classification and less misclassification 
of pulsars are reported in Table \ref{tab:covariate-E-combination} for training with datasets C and D. 
For balanced dataset, the true and false positive rates of the classifier increase significantly as compared to the imbalanced dataset. 
This bias can be attributed to the artificial data balancing using SMOTE which makes the classifier overconfident in predicting the pulsar candidates. 
Therefore, predictions with the imbalanced dataset are more reasonable than the balanced dataset. 
A threshold of 25\% on the percentage of true pulsar predictions out 
of total actual pulsars in the catalog and a maximum limit of 20\% of pulsar misclassification out of total pulsar candidates predicted in 
dataset E indicate that optimum results can be obtained with probability of pulsar prediction of 75\% using both RF and XGB for dataset C. 
Pulsar prediction probabilities of 95\% and 75\% with RF and XGB respectively yield similar results for dataset D. 
\par
In the recent TRAPUM L-band survey of pulsars \citep{Clark2023}, a total of 79 unassociated sources have been observed. Out of which, 9 new millisecond 
pulsars are discovered. All these sources have been appropriately categorised as Pulsar class in the 4FGL-DR4 catalog (version $\textit{gll\_psc\_v34.fit}$). 
The candidate source \textit{4FGL J1906.4-1757}, with radio pulsation detected, has not been classified as pulsar in the 4FGL-DR4 catalog. 
However, this source is predicted as Pulsar with probabilities of 92\% and 98\% by RF and XGB respectively in the present work. Among the 60 undetected 
sources in the survey (still under unassociated category in the 4FGL-DR4 catalog), about 80\% and 76\%  have been classified as likely to be pulsar by RF 
and XGB resepctively trained with balanced data (dataset B). However, combination of RF and XGB, as discussed above, can predict more than 55\% of undetected 
sources in the survey as likely to be pulsars. Similar constraints on prediction probabilites of more than 2200 unassociated sources, as discussed in 
Section \ref{classification}, suggest that the predicted number of pulsars is 173 when probability constraint of  75\% is set for both RF and XGB trained 
with original data (dataset A) and 151 when probability constraint of 95\% and 75\% is set for both RF and XGB  respectively trained with balanced 
data (dataset B).More recently, the Five-hundred-meter Aperture Spherical radio Telescope (FAST) has discoverd five new radio pulsars coincident 
with previously unassociated 4FGL-DR4 sources \citep{Fast2025A,Fast2025B}. These sources have also been predicted as pulsar candidates in the present 
work as summarized in Table \ref{tab:FAST-Discoveries}.
\begin{table}
	\caption{Summary of results from the actual performance evaluation of RF and XGB classifiers, trained with imbalanced dataset C and 
	balanced dataset D, for test dataset E.  N1 and N2 respectively represent the number of sources identified as highly likely  (probability $>$ 95\%) 
	and likely (probability $>$ 75\%) of the given class.}
\begin{center}
\begin{tabular}{p{1.5cm} p{1.5cm} p{2cm} p{2.5cm} p{2cm} p{2.5cm} p{2cm}}
\hline
Classifier &Dataset &True Pulsar &True Active Galactic Nuclei &False Pulsar &False Active Galactic Nuclei & Balanced Accuracy\\
		     &&(N1,N2)	 &(N1,N2)		     &(N1,N2)	     &(N1,N2) & (\%,\%)\\
\hline
	RF  &C  & 6, 24  &1484, 1623 &0, 5  &7, 22  & 48.8,65.6 \\
	RF  &D  & 20, 33 &1466, 1597 &4, 15 &5, 15 & 58.1,71.2\\
	XGB &C  & 15, 31 &1606, 1634 &2, 14 &18, 26  & 58.8,70.9\\
	XGB &D  & 31, 36 &1616, 1633 &13, 22 &22, 30 & 70.3,74.4\\
	\hline
\end{tabular}
\label{tab:covariate-E}
\end{center}
\end{table}

\begin{table}
	\caption{Summary of results from the actual performance evaluation for combination of RF and XGB classifiers, trained with imbalanced 
	dataset C and balanced dataset D, for test dataset E. $\mathrm P_{RF}$ and $\mathrm P_{XGB}$ are the probabilities of an unassociated source 
	being predicted as Pulsar by RF and XGB models respectively.}
\begin{center}
\begin{tabular}{c c c c c c}
\hline
$\mathrm{P_{RF}, P_{XGB}}$ & Dataset &Pulsar Predicted &True Pulsar &False Pulsar & F1-score\\
\hline
0.75, 0.75  & C & 29 & 24 & 5   & 0.479 \\
0.75, 0.95  & C & 17 & 15 & 2   & 0.339 \\
0.95, 0.75  & C & 6 & 6 & 0     &0.155\\
0.95, 0.95  & C & 6 & 6 & 0     &0.155\\
0.75, 0.75  & D & 46 & 32 & 14  &0.546 \\
0.75, 0.95  & D & 43 & 30 & 13  &0.525 \\
0.95, 0.75  & D & 24 & 20 & 4   &0.420\\
0.95, 0.95  & D & 24 & 20 & 4   &0.420\\
	\hline
\end{tabular}
\label{tab:covariate-E-combination}
\end{center}
\end{table}

\begin{table}
	\caption{Results on the prediction of five unassociated sources (recently detected as pulsars by FAST) as pulsar candiadtes 
	using RF and XGB classifiers when trained with balanced (SMOTE) and imbalanced datasets.}
\begin{center}
\begin{tabular}{c c c c c}
\hline
Source Name & $\mathrm {P_{RF}}$ &$\mathrm{P_{XGB}}$ &$\mathrm{P_{RF}(SMOTE)}$ &$\mathrm{P_{XGB}(SMOTE)}$\\
\hline
4FGL J0237.8+5238 & 0.966 & 0.962 & 0.994 & 0.999 \\
4FGL J0533.6+5945 & 0.842 & 0.901 & 0.977 & 0.999 \\
4FGL J1730.4-0359 & 0.969 & 0.956 & 0.994 & 0.999 \\
4FGL J1827.5+1141 & 0.792 & 0.890 & 0.980 & 0.983 \\
4FGL J1904.7-0708 & 0.718 & 0.836 & 0.968 & 0.989 \\
	\hline
\end{tabular}
\label{tab:FAST-Discoveries}
\end{center}
\end{table}

\section{Conclusions}\label{conclusions}
In this work, we have established two RF and XGB based binary classifiers for the classification of 2257 unassociated sources in the latest 
4FGL-DR4 catalog of the high energy gamma-ray sources released by the \emph{Fermi}-LAT collaboration. Results derived in this work indicate 
that large fraction of gamma-ray pulsar candidates is included in the existing population of the unassociated sources. The important findings 
of this work are summarized below:
\begin{itemize}
	\item Among more than 150 features available from observations, 8 features are identified to play crucial role in the classification 
		of the unassociated sources into Pulsar and Active Galactic Nuclei subclasses under the RFE framework. The spectral curvature 
		parameter ($\beta_{LP}$) has the highest importance in the classification. The covariate shift analysis suggests that balanced 
		accuracy does not improve further after including additional 4 weak features.

	\item  Under the covariate shift, the balanced accuracy of both the classifiers is found to be $\sim$ 70\% for the 
		test dataset E consisting of newly identified faint sources, which better represents the unassociated sources.    

       \item  Balanced accuracy under covariate shift analysis improves by $\sim 5\%$ when both the classifiers are trained with balanced 
	       dataset. However, the performance of XGB is found to be better than that of RF under both balanced and imbalanced training.
    
	\item RF classifier identifies 619 sources as likely pulsar candidates with an association probability of more than 75\%. Out of 619 
		candidates, 159 are predicted to be potential pulsar candidates with an association probability of higher than 95\%.

	\item In case of XGB classifier, 596 sources are classified as likely to be pulsars. And, the number of highly likely to be pulsar candidates 
		is found to be more than 334.
		
	\item Artificial data balancing using SMOTE helps in improving the F1-score of both the classifiers, when used in combination. 
		However, it increases the false positive rate significantly which is reflected in the increased number of pulsar predictions. 
		In order to get a realistic estimate of 
	      true pulsar predictions, the binary classification results based on imbalanced training dataset seem to be reasonable for targeted follow up 
		observations. However, for wide field surveys or initial screening, results based on the balanced training are crucial for determining the 
		hotspots in the universe for pulsar searches.

	\item Evaluation of the actual performance of RF and XGB classifiers indicates that XGB predicts more true pulsar candidates as likely and 
		highly likley to be pulsar candidates than RF. 

	\item The number of sources predicted as pulsars, among the unassociated sources in 4FGL-DR4 catalog, turns out to be 173 with an association 
		probability of 75\% when RF and XGB are used in combination for the imbalanced train data. For the balanced training data, this number is 151 
		with an association probability of 95\% and 75\% for RF and XGB respectively.
	\end{itemize}
These predictions are pivotal for the rapidly developing field of pulsar astronomy using the state-of-the-art ground-based gamma-ray telescopes like MAGIC, H.E.S.S., 
VERITAS, and MACE operating around the globe and the future CTA observatory. A dedicated monitoring of these potential candidates through pointed observations 
in the lower energy bands will provide additional informations to further quantify their exact nature. Apart from increasing the statistics of known gamma-ray 
pulsars, results reported in this work are important for population syntheses of the normal and millisecond pulsars. This also underlines importance of ongoing 
investigations to understand the physics of high energy gamma-ray emission from pulsars. 
\section*{Declaration of Competing Interest}
The authors declare that they have no known competing financial interests or personal relationships that could have appeared to influence the work reported in this paper.
\section*{Acknowledgements}
We thank the two anonymous reviewers for their very insightful constructive comments which have greatly helped in improving the contents of 
the manuscript scientifically. This work made extensive use of the ATNF pulsar catalog (Manchester et al. 2005). We would like to thank 
the Fermi Science Support Center (FSSC) for the public availability of data. 
\section*{Data Availability}
The results of this study are available in machine readable format (.csv) as supplementary material online.
\bibliography{MS}

\end{document}